\documentclass[12pt,prd,showpacs,tightenlines,nofootinbib]{revtex4}
\usepackage{bm}
\usepackage{graphics}
\usepackage{rotating}
\usepackage{epsfig}
\begin{document}
\title{\begin{flushright}{\rm\normalsize HU-EP-05/18}\end{flushright}
Masses of heavy baryons in the relativistic quark model}
\author{D. Ebert}
\affiliation{Institut f\"ur Physik, Humboldt--Universit\"at zu Berlin,
Newtonstr. 15, D-12489  Berlin, Germany}
\author{R. N. Faustov}
\author{V. O. Galkin}
\affiliation{Institut f\"ur Physik, Humboldt--Universit\"at zu Berlin,
Newtonstr. 15, D-12489 Berlin, Germany}
\affiliation{Dorodnicyn Computing Centre, Russian Academy of Sciences,
  Vavilov Str. 40, 119991 Moscow, Russia}

\begin{abstract}
The masses of the ground state heavy baryons consisting of two light
($u,d,s$) and one heavy ($c,b$) quarks are calculated in the 
heavy-quark--light-diquark approximation within the constituent quark
model. The light quarks, forming the diquark, and the light diquark in the
baryon are treated completely relativistically. The expansion in
$v/c$ up to the second order  is
used only for the heavy ($b$ and $c$) quarks. 
The diquark-gluon interaction is taken modified by the form
factor describing the light diquark structure in terms of the diquark
wave functions. An overall
reasonable agreement of the obtained predictions with available
experimental data and previous theoretical results is found.       
 
\end{abstract}


\maketitle
\section{Introduction}

The description of baryons within the constituent quark models is a
very important problem in quantum chromodynamics (QCD). Since the
baryon is a three-body system, its theory is much more complicated
compared to the two-body meson system.  The quark-diquark picture of a
baryon \cite{apefl,erv} is the popular approximation widely used to
describe the baryon properties \cite{apefl,erv,kkp,jaffe,wil}. The methods
of heavy quark effective theory (HQET) proved to be very fruitful in
predicting  the
properties of the heavy-light $q\bar Q$ mesons ($B$ and $D$). This success
suggests to apply these methods to heavy-light baryons. 
In our paper \cite{efgm} we considered the simplest
baryonic systems of this kind, the doubly heavy baryons
($qQQ$). The two heavy quarks
($b$ or $c$) compose in this case a bound diquark system in the
antitriplet colour state which serves as a localized colour
source. The light quark is orbiting around it and the resulting
effective two-body system strongly resembles the heavy-light $B$ and
$D$ mesons. The main distinction is that the quark-diquark interaction
is not point-like due to the heavy-diquark form factor
\cite{efgm}.  The heavy-quark expansion in $1/m_Q$  can be
used here, and the light quark is treated fully relativistically.

The experience acquired in the investigation of doubly-heavy baryons
and the recent success in the relativistic description of light mesons
\cite{lmes} made it possible to study the baryons with one heavy quark
($b$ or $c$), too. In this case we assume that the
heavy-quark--light-diquark configuration dominates. Thus the
three-body problem is again reduced to the two-body
one. A crucial assumption of this quark-diquark picture consists in
neglecting the influence of the third quark on the internal diquark
dynamics. In the model with the harmonic-oscillator pair
interactions, for instance, this effect increases the interaction
strength by a factor 3/2, and thus a significant increase by a factor
$\sqrt{3/2}$ of the internal excitation energies of the diquark is
achieved. It is hoped that for linear confinement and relativistic
kinematics this effect would be modified and become smaller, but the
answer could come only from 
a thorough treatment of the relativistic three-body problem and
comparison with the quark-diquark approximation. 
In the considered version of the quark-diquark picture such influence
could be partially reproduced by the account of the heavy-light
diquark. We assume that such contributions are small and thus
the heavy quark influence on the diquark dynamics is also
small. Such assumption is necessary to preserve the presumed universal
nature of the diquark \cite{jaffe}. Otherwise the diquark properties
would be very different in such hadronic systems as baryons,
tetraquarks, pentaquarks, etc.   
Unfortunately, the $1/m_Q$ expansion cannot be reliably applied
for the $c$ quark since its mass proves to be comparable with the mass
of the light diquark. So we use instead the $v/c$ expansion for the
heavy quark and a completely relativistic description for the light
quarks.  Fortunately, our predictions can be compared with the rather
large amount of experimental data for the ground state baryons with
one heavy quark (mainly the $c$).

\section{Relativistic quark model}
\label{sec:rqm}

In the quasipotential approach and quark-diquark picture of
heavy baryons the interaction of two light quarks in a diquark and the heavy
quark interaction with a light diquark in a baryon are described by the
diquark wave function ($\Psi_{d}$) of the bound quark-quark state
and by the baryon wave function ($\Psi_{B}$) of the bound quark-diquark
state respectively,  which satisfy the
quasipotential equation \cite{3} of the Schr\"odinger type \cite{4}
\begin{equation}
\label{quas}
{\left(\frac{b^2(M)}{2\mu_{R}}-\frac{{\bf
p}^2}{2\mu_{R}}\right)\Psi_{d,B}({\bf p})} =\int\frac{d^3 q}{(2\pi)^3}
 V({\bf p,q};M)\Psi_{d,B}({\bf q}),
\end{equation}
where the relativistic reduced mass is
\begin{equation}
\mu_{R}=\frac{E_1E_2}{E_1+E_2}=\frac{M^4-(m^2_1-m^2_2)^2}{4M^3},
\end{equation}
and $E_1$, $E_2$ are given by
\begin{equation}
\label{ee}
E_1=\frac{M^2-m_2^2+m_1^2}{2M}, \quad E_2=\frac{M^2-m_1^2+m_2^2}{2M},
\end{equation}
here $M=E_1+E_2$ is the bound state mass (diquark or baryon),
$m_{1,2}$ are the masses of light quarks ($q_1$ and $q_2$) which form
the diquark or of the light diquark ($d$) and heavy quark ($Q$) which form
the heavy baryon ($B$), and ${\bf p}$  is their relative momentum.  
In the center of mass system the relative momentum squared on mass shell 
reads
\begin{equation}
{b^2(M) }
=\frac{[M^2-(m_1+m_2)^2][M^2-(m_1-m_2)^2]}{4M^2}.
\end{equation}

The kernel 
$V({\bf p,q};M)$ in Eq.~(\ref{quas}) is the quasipotential operator of
the quark-quark or quark-diquark interaction. It is constructed with
the help of the
off-mass-shell scattering amplitude, projected onto the positive
energy states. In the following analysis we closely follow the
similar construction of the quark-antiquark interaction in mesons
which were extensively studied in our relativistic quark model
\cite{egf}. For
the quark-quark interaction in a diquark we use the relation
$V_{qq}=V_{q\bar q}/2$ arising under the assumption about the octet
structure of the interaction  from the difference of the $qq$ and
$q\bar q$  colour states. An important role in this construction is
played by the Lorentz-structure of the confining  interaction. 
In our analysis of mesons while  
constructing the quasipotential of the quark-antiquark interaction, 
we adopted that the effective
interaction is the sum of the usual one-gluon exchange term with the mixture
of long-range vector and scalar linear confining potentials, where
the vector confining potential contains the Pauli terms.  
We use the same conventions for the construction of the quark-quark
and quark-diquark interactions in the baryon. The
quasipotential  is then defined by \cite{efgm,egf} 

(a) for the quark-quark ($qq$) interaction
 \begin{equation}
\label{qpot}
V({\bf p,q};M)=\bar{u}_{1}(p)\bar{u}_{2}(-p){\cal V}({\bf p}, {\bf
q};M)u_{1}(q)u_{2}(-q),
\end{equation}
with
\[
{\cal V}({\bf p,q};M)=\frac12\left[\frac43\alpha_sD_{ \mu\nu}({\bf
k})\gamma_1^{\mu}\gamma_2^{\nu}+ V^V_{\rm conf}({\bf k})
\Gamma_1^{\mu}({\bf k})\Gamma_{2;\mu}(-{\bf k})+
 V^S_{\rm conf}({\bf k})\right],
\]

(b) for quark-diquark ($Qd$) interaction
\begin{eqnarray}
\label{dpot}
V({\bf p,q};M)&=&\frac{\langle d(P)|J_{\mu}|d(Q)\rangle}
{2\sqrt{E_d(p)E_d(q)}} \bar{u}_{Q}(p)  
\frac43\alpha_SD_{ \mu\nu}({\bf 
k})\gamma^{\nu}u_{Q}(q)\cr
&&+\psi^*_d(P)\bar u_Q(p)J_{d;\mu}\Gamma_Q^\mu({\bf k})
V_{\rm conf}^V({\bf k})u_{Q}(q)\psi_d(Q)\cr 
&&+\psi^*_d(P)
\bar{u}_{Q}(p)V^S_{\rm conf}({\bf k})u_{Q}(q)\psi_d(Q), 
\end{eqnarray}
where $\alpha_s$ is the QCD coupling constant, $\langle
d(P)|J_{\mu}|d(Q)\rangle$ is the vertex of the 
diquark-gluon interaction which is discussed in detail
below $\Big[$$P=(E_d,-{\bf p})$ and $Q=(E_d,-{\bf q})$,
$E_d=(M^2-m_Q^2+M_d^2)/(2M)$ $\Big]$. $D_{\mu\nu}$ is the 
gluon propagator in the Coulomb gauge
\begin{equation}
D^{00}({\bf k})=-\frac{4\pi}{{\bf k}^2}, \quad D^{ij}({\bf k})=
-\frac{4\pi}{k^2}\left(\delta^{ij}-\frac{k^ik^j}{{\bf k}^2}\right),
\quad D^{0i}=D^{i0}=0,
\end{equation}
and ${\bf k=p-q}$; $\gamma_{\mu}$ and $u(p)$ are 
the Dirac matrices and spinors
\begin{equation}
\label{spinor}
u^\lambda({p})=\sqrt{\frac{\epsilon(p)+m}{2\epsilon(p)}}
\left(\begin{array}{c}
1\\ \displaystyle\frac{\mathstrut\bm{\sigma}{\bf p}}
{\mathstrut\epsilon(p)+m}
\end{array}\right)
\chi^\lambda,
\end{equation}
with $\epsilon(p)=\sqrt{{\bf p}^2+m^2}$. 

The diquark state in the confining part of the quark-diquark
quasipotential (\ref{dpot}) is described by the wave functions
\begin{equation}
  \label{eq:ps}
  \psi_d(p)=\left\{\begin{array}{ll}1 &\qquad \text{ for scalar diquark}\\
\varepsilon_d(p) &\qquad \text{ for axial vector diquark}
\end{array}\right. ,
\end{equation}
where the four vector
\begin{equation}\label{pv}
\varepsilon_d(p)=\left(\frac{(\bm{\varepsilon}_d {\bf
p})}{M_d},\bm{\varepsilon}_d+ \frac{(\bm{\varepsilon}_d {\bf p}){\bf
  p}}{M_d(E_d(p)+M_d)}\right), \qquad \varepsilon^\mu_d(p) p_\mu=0,  
\end{equation} 
is the polarization vector of the axial vector
diquark with momentum ${\bf p}$, $E_d(p)=\sqrt{{\bf p}^2+M_d^2}$ and
$\varepsilon_d(0)=(0,\bm{\varepsilon}_d)$ is the polarization vector in
the diquark rest frame. The effective long-range vector vertex of the
diquark can be presented in the form  
\begin{equation}
  \label{eq:jc}
  J_{d;\mu}=\left\{\begin{array}{ll}
  \frac{\displaystyle (P+Q)_\mu}{\displaystyle
  2\sqrt{E_d(p)E_d(q)}}&\qquad \text{ for scalar diquark}\cr
\frac{\displaystyle (P+Q)_\mu}{\displaystyle2\sqrt{E_d(p)E_d(q)}}
  -\frac{\displaystyle i\mu_d}{\displaystyle 2M_d}\Sigma_\mu^\nu 
\tilde k_\nu
  &\qquad \text{ for axial 
  vector diquark}\end{array}\right. ,
\end{equation}
where $\tilde k=(0,{\bf k})$. Here the antisymmetric tensor
\begin{equation}
  \label{eq:Sig}
  \left(\Sigma_{\rho\sigma}\right)_\mu^\nu=-i(g_{\mu\rho}\delta^\nu_\sigma
  -g_{\mu\sigma}\delta^\nu_\rho)
\end{equation}
and the axial vector diquark spin ${\bf S}_d$ is given by
$(S_{d;k})_{il}=-i\varepsilon_{kil}$. We choose the total
chromomagnetic moment of the axial vector 
diquark $\mu_d=2$ \cite{efgm}.

The effective long-range vector vertex of the quark is
defined by \cite{egf,sch}
\begin{equation}
\Gamma_{\mu}({\bf k})=\gamma_{\mu}+
\frac{i\kappa}{2m}\sigma_{\mu\nu}\tilde k^{\nu}, \qquad \tilde
k=(0,{\bf k}),
\end{equation}
where $\kappa$ is the Pauli interaction constant characterizing the
anomalous chromomagnetic moment of quarks. In the configuration space
the vector and scalar confining potentials in the nonrelativistic
limit reduce to
\begin{eqnarray}
V^V_{\rm conf}(r)&=&(1-\varepsilon)V_{\rm conf}(r),\nonumber\\
V^S_{\rm conf}(r)& =&\varepsilon V_{\rm conf}(r),
\end{eqnarray}
with 
\begin{equation}
V_{\rm conf}(r)=V^S_{\rm conf}(r)+
V^V_{\rm conf}(r)=Ar+B,
\end{equation}
where $\varepsilon$ is the mixing coefficient.

The constituent quark masses $m_b=4.88$ GeV, $m_c=1.55$ GeV,
$m_u=m_d=0.33$ GeV, $m_s=0.5$ GeV and 
the parameters of the linear potential $A=0.18$ GeV$^2$ and $B=-0.3$ GeV
have the usual values of quark models.  The value of the mixing
coefficient of vector and scalar confining potentials $\varepsilon=-1$
has been determined from the consideration of charmonium radiative
decays \cite{efg} and the heavy quark expansion \cite{fg}. 
Finally, the universal Pauli interaction constant $\kappa=-1$ has been
fixed from the analysis of the fine splitting of heavy quarkonia ${
}^3P_J$- states \cite{efg}. In the literature it is widely discussed
the 't~Hooft-like interaction between quarks induced by instantons \cite{dk}.
This interaction can be partly described by introducing the quark
anomalous chromomagnetic moment having an approximate value
$\kappa=-0.744$ (Diakonov \cite{dk}). This value is of the same
sign and order of magnitude as the Pauli constant $\kappa=-1$ in our
model. Thus the Pauli term incorporates at least some part of the
instanton contribution to the $q\bar q$ interaction.  Note that the 
long-range chromomagnetic contribution to the potential in our model
is proportional to $(1+\kappa)$ and thus vanishes for the 
chosen value of $\kappa=-1$.

\section{Properties of light diquarks}
\label{sec:pld}

At a first step, we calculate the masses and form factors of the light
diquark. As it is well known, the light quarks are highly
relativistic, which makes the $v/c$ expansion inapplicable and thus,
a completely relativistic treatment is required. To achieve this goal in
describing light 
diquarks, we closely follow our recent consideration of the spectra of light
mesons \cite{lmes} and adopt the same procedure to make the relativistic
quark potential local by replacing
$\epsilon_{1,2}(p)\equiv\sqrt{m_{1,2}^2+{\bf p}^2}\to E_{1,2}$  
(see discussion in Ref.~\cite{lmes}).  As a result,
the light quark-quark interaction 
(\ref{qpot})  in the diquark state, which is 1/2  of the $q\bar q$
interaction in light mesons, consists of the sum of the 
spin-independent and spin-dependent parts \cite{lmes}
\begin{equation}
  \label{eq:v}
  V(r)= V_{\rm SI}(r)+ V_{\rm SD}(r),
\end{equation}
where the spin-independent potential for the $S$-wave states (${\bf
  L}^2=0$) has the form 
\begin{eqnarray}
  \label{eq:vsi}
  V_{\rm SI}(r)&=&\frac12\Biggl[V_{\rm Coul}(r)+V_{\rm conf}(r)+
\frac{(E_1^2-m_1^2+E_2^2-m_2^2)^2}{4(E_1+m_1)(E_2+m_2)}\Biggl\{
\frac1{E_1E_2}V_{\rm Coul}(r)\cr
&& +\frac1{m_1m_2}\Biggl(1+(1+\kappa)\Biggl[(1+\kappa)\frac{(E_1+m_1)(E_2+m_2)}
{E_1E_2}\cr
&&-\left(\frac{E_1+m_1}{E_1}+\frac{E_1+m_2}{E_2}\right)\Biggr]\Biggr)
V^V_{\rm conf}(r)
+\frac1{m_1m_2}V^S_{\rm conf}(r)\Biggr\}\cr
&&+\frac14\left(\frac1{E_1(E_1+m_1)}\Delta
\tilde V^{(1)}_{\rm Coul}(r)+\frac1{E_2(E_2+m_2)}\Delta
\tilde V^{(2)}_{\rm Coul}(r)\right)\cr
&&-\frac14\left[\frac1{m_1(E_1+m_1)}+\frac1{m_2(E_2+m_2)}-(1+\kappa)
\left(\frac1{E_1m_1}+\frac1{E_2m_2}\right)\right]\Delta V^V_{\rm
conf}(r)\cr
&&+\frac{(E_1^2-m_1^2+E_2^2-m_2^2)}{8m_1m_2(E_1+m_1)(E_2+m_2)} 
\Delta V^S_{\rm conf}(r)\Biggr], 
\end{eqnarray}
and the spin-dependent potential is given by
\begin{eqnarray}
  \label{eq:vsd}
   V_{\rm SD}(r)&=&\frac1{3E_1E_2}\Biggl[\Delta \bar V_{\rm Coul}(r)
+\left(\frac{E_1-m_1}{2m_1}-(1+\kappa)\frac{E_1+m_1}{2m_1}\right)\cr
&&\qquad\quad\times
\left(\frac{E_2-m_2}{2m_2}-(1+\kappa)\frac{E_2+m_2}{2m_2}\right)
\Delta V^V_{\rm conf}(r)\Biggr]{\bf S}_1{\bf S}_2,
\end{eqnarray}
with \cite{lmes,egf}
\begin{eqnarray}
  \label{eq:tv}
V_{\rm Coul}(r)&=&-\frac43\frac{\alpha_s}{r},\cr
\tilde V^{(i)}_{\rm Coul}(r)&=&V_{\rm Coul}(r)\frac1{\displaystyle\left(1+
\eta_i\frac43\frac{\alpha_s}{E_i}\frac1{r}\right)\left(1+
\eta_i\frac43\frac{\alpha_s}{E_i+m_i}\frac1{r}\right)},\qquad (i=1,2),\cr
  \bar V_{\rm Coul}(r)&=&V_{\rm Coul}(r)\frac1{\displaystyle\left(1+
\eta_1\frac43\frac{\alpha_s}{E_1}\frac1{r}\right)\left(1+
\eta_2\frac43\frac{\alpha_s}{E_2}\frac1{r}\right)}, 
\qquad \eta_{1,2}=\frac{m_{2,1}}{m_1+m_2}.
\end{eqnarray}
Here we put  $\alpha_s\equiv\alpha_s(\mu_{12}^2)$ with $\mu_{12}=2m_1
m_2/(m_1+m_2)$ and use for $\alpha_s(\mu^2)$ the
simplest model with freezing \cite{bvb}
\begin{equation}
  \label{eq:alpha}
  \alpha_s(\mu^2)=\frac{4\pi}{\displaystyle\beta_0
\ln\frac{\mu^2+M_0^2}{\Lambda^2}}, \qquad \beta_0=11-\frac23n_f,
\end{equation}
where the background mass is $M_0=2.24\sqrt{A}=0.95$~GeV \cite{bvb} and
$\Lambda=413$~MeV was fixed from fitting the $\rho$ mass. We put
 the number of flavours $n_f=2$ for $ud$, 
$us$ diquarks and $n_f=3$ for $ss$ diquark, cf. \cite{lmes} . As a
result we obtain  
$\alpha_s(\mu_{ud}^2)=0.730$, $\alpha_s(\mu_{us}^2)=0.711$ and
$\alpha_s(\mu_{ss}^2)=0.731$.   

The quasipotential equation (\ref{quas}) is solved numerically for the
complete relativistic potential (\ref{eq:v}) which depends on the
diquark mass in a complicated highly nonlinear way.  The obtained
ground state masses of scalar  and axial
vector  light diquarks are presented in
Table~\ref{tab:mass}. These  masses are in good agreement with values
found in Ref.~\cite{efkr} within the Nambu--Jona-Lasinio model.
It follows from  Table~\ref{tab:mass} that the mass difference between
the scalar and vector diquark decreases from $\sim 200$ to $\sim 120$ MeV,
when one of the $u,d$ quarks is replaced by the $s$ quark in accord
with the statement of Ref.~\cite{jaffe}.    
\begin{table}
  \caption{Masses of light ground state diquarks (in MeV). S and A
    denotes scalar and axial vector diquarks antisymmetric $[q,q']$ and
    symmetric $\{q,q'\}$ in flavour, respectively. }
  \label{tab:mass}
\begin{ruledtabular}
\begin{tabular}{cccc}
Quark& Diquark&  
\multicolumn{2}{l}{\underline{\hspace{2.9cm}Mass\hspace{2.9cm}}}
\hspace{-3.6cm} \\
content &type &this work& Ref.\cite{efkr}$^*$\\
\hline
$[u,d]$& S & 710 & 705 \\
$\{u,d\}$& A & 909 & 875 \\
$[u,s]$ & S& 948 & 895 \\
$\{u,s\}$& A & 1069 & 1050 \\
$\{s,s\}$& A & 1203 & 1215 
  \end{tabular}
\end{ruledtabular}
\flushleft{${}^*$ For $G_1/G_{\rm meson}=1.1$}
\end{table}
 
In order to determine the diquark interaction with the gluon field, which
takes into account the diquark structure, it is
necessary to calculate the corresponding matrix element of the quark
current between diquark states. This diagonal matrix element can be
parametrized by the following set of elastic form factors

(a) scalar diquark ($S$)
\begin{equation}
  \label{eq:sff}
  \langle S(P)\vert J_\mu \vert S(Q)\rangle=h_+(k^2)(P+Q)_\mu,
\end{equation}

(b) axial vector diquark ($A$) 
\begin{eqnarray}
  \label{eq:avff}
\langle A(P)\vert J_\mu \vert A(Q)\rangle&=&
-[\varepsilon_d^*(P)\cdot\varepsilon_d(Q)]h_1(k^2)(P+Q)_\mu\cr
&&+h_2(k^2)
\left\{[\varepsilon_d^*(P) \cdot Q]\varepsilon_{d;\mu}(Q)+
  [\varepsilon_d(Q) \cdot P] 
\varepsilon^*_{d;\mu}(P)\right\}\cr
&&+h_3(k^2)\frac1{M_{A}^2}[\varepsilon^*_d(P) \cdot Q]
    [\varepsilon_d(Q) \cdot P](P+Q)_\mu, 
\end{eqnarray}
where $k=P-Q$ and $\varepsilon_d(P)$ is the polarization vector of the
axial vector diquark (\ref{pv}). 

\begin{figure}
\centerline{\includegraphics{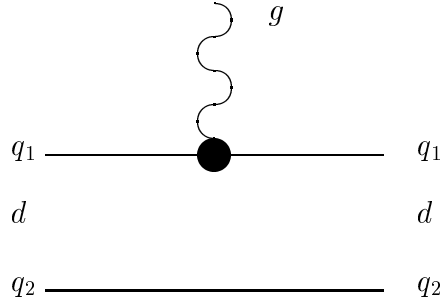}}
  \caption{\label{fig:pic1} The vertex function $\Gamma$ in the impulse
    approximation. The gluon interaction only with one
light quark is shown.}
\end{figure}

In the quasipotential approach,  the
matrix element of the quark current $J_\mu=\bar q
\gamma^\mu q$  between the diquark states ($d$) has the form \cite{f}
\begin{equation}\label{mxet}
\langle d(P) \vert J_\mu (0) \vert d(Q)\rangle
=\int \frac{d^3p\, d^3q}{(2\pi )^6} \bar \Psi^{d}_P({\bf
p})\Gamma _\mu ({\bf p},{\bf q})\Psi^{d}_Q({\bf q}),\end{equation}
where $\Gamma _\mu ({\bf p},{\bf
q})$ is the two-particle vertex function and  $\Psi^{d}_P$ are the
diquark wave functions projected onto the positive energy states of
quarks and boosted to the moving reference frame with momentum $P$.
In the impulse approximation the vertex function $\Gamma$ is shown in
Fig.~\ref{fig:pic1}.  The corresponding vertex
function is given by
\begin{equation}\label{gam1}
\Gamma_\mu ^{(1)}({\bf p},{\bf q})=\bar
u_{q_1}(p_1)\gamma^\mu u_{q_1}(q_1)(2\pi)^3\delta({\bf p}_2-{\bf q}_2)+
(1\leftrightarrow 2),
\end{equation} 
where \cite{f} 
\begin{eqnarray*}\label{pqm}
p_{1,2}&=&\epsilon_{1,2}(p)\frac{P}{{\cal M}_d}\pm\sum_{i=1}^3n^{(i)}(P)p^i,
\quad {\cal M}_d=\epsilon_1(p)+\epsilon_2(p),\cr 
q_{1,2}&=&\epsilon_{1,2}(q)\frac{Q}{{\cal M'}_d}\pm\sum_{i=1}^3n^{(i)}(Q)q^i,
\quad {\cal M'}_d=\epsilon_1(q)+\epsilon_2(q),
\end{eqnarray*}
and $n^{(i)}$ are three four vectors defined by
$$ n^{(i)\mu}(P)=\left\{ \frac{P^i}{{\cal M}_d},\ \delta_{ij}+
\frac{P^iP^j}{{\cal M}_d[E_d(P)+{\cal M}_d]}\right\}, 
\quad E_d(P)=\sqrt{{\bf P}^2+{\cal M}_d^2}.$$
After making necessary computations,
the expression for $\Gamma$ should be continued in
${\cal M}_d$ and ${\cal M'}_d$  to the  diquark mass $M_d$.

\begin{figure}
\includegraphics{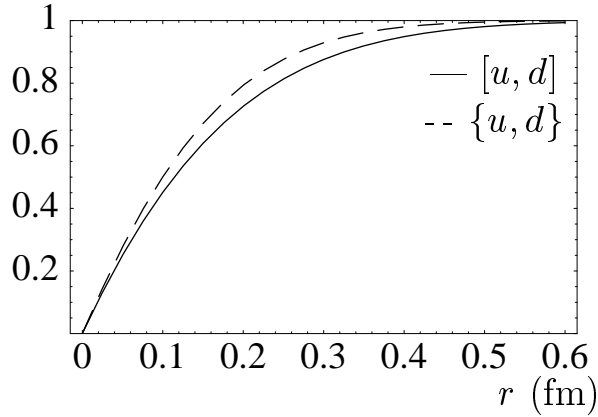}
\caption{\label{fig:ff} The form factors $F(r)$ for the scalar
$[u,d]$ (solid line) and axial vector $\{u,d\}$ (dashed line) diquarks.}  
\end{figure}

Substituting the vertex function $\Gamma^{(1)}$ given by
Eq.~(\ref{gam1}) in the matrix element (\ref{mxet}) and  comparing the
resulting expressions with 
the form factor decompositions (\ref{eq:sff}) and (\ref{eq:avff}), we
find 
\begin{eqnarray*}
  h_+(k^2)&=&h_1(k^2)=h_2(k^2)=F({\bf k}^2),\cr
h_3(k^2)&=&0,
\end{eqnarray*}
\begin{eqnarray}\label{eq:hf}
F({\bf k}^2)&=&\frac{\sqrt{E_{d}M_{d}}}{E_{d}+M_{d}}
  \int \frac{d^3p}{(2\pi )^3} \bar\Psi_{d}
\left({\bf p}+
\frac{2\epsilon_{2}(p)}{E_{d}+M_{d}}{\bf k } \right)
\sqrt{\frac{\epsilon_1(p)+m_1}{\epsilon_1(p+k)+m_1}}
\Biggl[\frac{\epsilon_1(p+k)+\epsilon_1(p)}
{2\sqrt{\epsilon_1(p+k)\epsilon_1(p)}}\cr
&&+\frac{\bf p k}{2\sqrt{\epsilon_1(p+k)\epsilon_1(p)}
(\epsilon_1(p)+m_1)} \Biggr]\Psi_{d}({\bf
  p})+(1\leftrightarrow 2),
\end{eqnarray}
where $\Psi_{d}$ are the diquark wave functions.
We calculated the corresponding form factors $F(r)/r$ which are the Fourier
transforms of $F({\bf k}^2)/{\bf k}^2$ using the diquark wave
functions found 
by numerical solving the quasipotential equation. In Fig.~\ref{fig:ff}
the functions $F(r)$ for the scalar $[u,d]$ and axial vector $\{u,d\}$
diquarks  are shown as an example.
Our estimates show that this form factor can be approximated  with a
high accuracy by the expression 
\begin{equation}
  \label{eq:fr}
  F(r)=1-e^{-\xi r -\zeta r^2},
\end{equation}
which agrees with previously used approximations \cite{efgm}.
The values of parameters $\xi$ and $\zeta$ for light
diquark scalar $[q,q']$ and axial vector $\{q,q'\}$ ground states are
given in Table~\ref{tab:fcc}. 
As we see, the functions $F(r)$ vanish at $r= 0$ and tend to
unity for large values of $r$. Such a behaviour can be easily understood
intuitively. At large distances a diquark can be well approximated by a
point-like object and its internal structure cannot be resolved. When
the distance to the diquark decreases the internal structure plays a more
important role. As the distance approaches zero, the interaction weakens and
turns to zero for $r=0$. Thus the function $F(r)$
gives an important contribution to the short-range part of the
interaction of the heavy quark with the light diquark in the baryon and
can be neglected for the long-range (confining) interaction.  

\begin{table}
\caption{\label{tab:fcc}Parameters  $\xi$ and $\zeta$ for ground state
  light diquarks.}
\begin{ruledtabular}
\begin{tabular}{cccc}
Quark &Diquark& $\xi$  & $\zeta$  \\
content& type&(GeV)&(GeV$^2$)\\
\hline
$[u,d]$&S & 1.09 & 0.185  \\
$\{u,d\}$&A &1.185 & 0.365  \\
$[u,s]$& S & 1.23 & 0.225 \\
$\{u,s\}$& A & 1.15 & 0.325\\
$\{s,s\}$ & A& 1.13 & 0.280
\end{tabular}
\end{ruledtabular}
\end{table}

\section{Masses of heavy baryons}
\label{sec:mhb}

At a second step, we calculate the masses of heavy baryons as the bound
state of a heavy quark and light diquark.
For  the potential of the heavy-quark--light diquark
interaction (\ref{dpot}) we use the expansion in $p/m_Q$. Since the
light diquark is not  heavy enough for the applicability of a $p/m_d$
expansion, it should be treated fully relativistically. To achieve
this goal  and simplify the potential we follow the same procedure,
which was used for light quarks in a diquark,  and replace
the diquark energies $E_d(p)\equiv\sqrt{{\bf p}^2+M_d^2}\to
E_d\equiv(M^2-m_Q^2+M_d^2)/(2M)$ in  
Eqs.~(\ref{dpot}), (\ref{eq:jc}). This substitution makes the Fourier
transform of the potential (\ref{dpot}) local.  
At leading order in $p/m_Q$ the resulting
potential   for the $S$-wave states (${\bf L}^2=0$, ${\bf LS}=0$) is
the same for scalar and axial vector 
diquarks and is given by
\begin{equation}
\label{v0}
 V^{(0)}(r)= \hat V_{\rm Coul}(r)
+V_{\rm conf}(r),
\end{equation}
$$\hat V_{\rm Coul}(r)=-\frac43\alpha_s\frac{ F(r)}{r}, \qquad 
V_{\rm   conf}(r)=Ar+B,$$
 where $\hat V_{\rm Coul}(r)$ is the smeared Coulomb
potential (which accounts for the diquark structure) and $\alpha_s$ is
given by Eq.~(\ref{eq:alpha}) with $N_f=3$. The masses of
baryons with spin 1/2 and 3/2, containing the axial vector diquark, are
degenerate in this approximation since the spin-spin
interaction arises only at first order in $p/m_{Q}$. Solving
Eq.~(\ref{quas}) numerically  we  get the spin-independent part of
the baryon wave function $\Psi_B$. Then the total baryon wave function 
is a product of $\Psi_B$ and the spin-dependent part $U_B$ (for
details see Eq. (43) of Ref.~\cite{dhbd}). 

The leading order degeneracy of heavy baryon states is broken by $p/m_{Q}$
corrections. 
The ground-state quark-diquark potential (\ref{dpot}) up to the second
order of the $p/m_{Q}$ expansion is given by the
following expressions: 

(a) scalar diquark
\begin{eqnarray}
\label{svcor}\!\!\!\!\!\!\!
\delta V(r)&=&\frac{1}{E_dm_{Q}}\Bigg\{{\bf p}\left[\hat V_{\rm
Coul}(r)+V^V_{\rm conf}(r)\right]{\bf p}
-\frac{1}{4}\Delta V^V_{\rm conf}(r)\Bigg\}\cr
&&+\frac1{m_Q^2}\left\{\frac18\Delta\left(\hat V_{\rm
Coul}(r)+V^S_{\rm conf}(r)-[1-2(1+\kappa)]V^V_{\rm conf}(r)
\right)
-\frac12{\bf p}V^S_{\rm conf}(r){\bf p}\right\},
\end{eqnarray}

(b) axial vector diquark
\begin{eqnarray}
\label{avcor}\!\!\!\!\!\!\!
\delta V(r)&=&\frac{1}{E_dm_{Q}}\Bigg\{{\bf p}\left[\hat V_{\rm
Coul}(r)+V^V_{\rm conf}(r)\right]{\bf p}
-\frac{1}{4}\Delta V^V_{\rm conf}(r)\cr
& & +\frac23\left[\Delta \hat V_{\rm Coul}(r)+(1+\kappa)\Delta V^V_{\rm 
conf}(r)\right]{\bf S}_{d}{\bf S}_Q\Bigg\}\cr
&&+\frac1{m_Q^2}\left\{\frac18\Delta\left(\hat V_{\rm
Coul}(r)+V^S_{\rm conf}(r)-[1-2(1+\kappa)]V^V_{\rm conf}(r)
\right)
-\frac12{\bf p}V^S_{\rm conf}(r){\bf p}\right\},
\end{eqnarray}
where  ${\bf S}_{d}$ and ${\bf S}_{Q}$ are the light diquark and heavy
quark spins, respectively.
It is necessary to note that the confining vector interaction gives a
contribution to the spin-dependent part which is proportional to
$(1+\kappa)$. Thus it vanishes for the chosen value
of $\kappa=-1$, while the confining vector contribution to 
the spin-independent part is nonzero. 

Now we can calculate the mass spectra of heavy 
baryons with the account of all  corrections  of
order $p^2/m_Q^2$. For this purpose we consider 
Eq.~(\ref{quas}) with  the quasipotential
which is the sum of the leading order potential $V^{(0)}(r)$  (\ref{v0}) and
the correction $\delta V(r)$ (\ref{svcor}) and (\ref{avcor}). We
multiply this equation from the left 
by the quasipotential wave function of a bound state and
integrate both sides over the relative momentum. Within the
adopted accuracy of calculations, we can use for the resulting matrix
elements the wave functions of Eq.~(\ref{quas}) with the leading order
potential $V^{(0)}(r)$.  
In this way we obtain the mass formula 
\begin{equation}
\label{mform}
\frac{b^2(M)}{2\mu_R}=\frac{\langle{\bf
p}^2\rangle}{2\mu_{R}}+\langle V^{(0)}(r)\rangle+\langle\delta V(r)  \rangle.
\end{equation}
The contribution of the spin-spin interaction in (\ref{avcor}) is
proportional to 
\begin{equation}
  \label{eq:ss}
  \langle {\bf S}_d {\bf S}_Q\rangle=\frac12\left[J(J+1)-S_d(S_d+1)
-\frac34\right],
\end{equation}
where ${\bf J}={\bf S}_d+{\bf S}_Q$ is the spin of the ground state
heavy baryon. \footnote{It should be mentioned that
  $Q[sq]\leftrightarrow Q\{sq\}$ mixing can exist in conventional
  constituent quark models with three quarks, but it is absent for
  ground states in our approach.}

The calculated values of the baryon masses are given in
Table~\ref{tab:bm} in comparison with some theoretical predictions
\cite{ci,rlp,sav,j,mlw} and experimental data \cite{pdg}. 
  
\begin{table}
\caption{\label{tab:bm} Masses of the ground state heavy baryons (in MeV).}
\begin{ruledtabular}
\begin{tabular}{ccccccccc}
Baryon &$I(J^P)$& 
\multicolumn{6}{l}{\underline{\hspace{4.9cm}Theory\hspace{4.9cm}}} 
\hspace{-4.1cm} 
& Experiment\\
& & this work &Ref.~\cite{ci}  &Ref.~\cite{rlp}&Ref.~\cite{sav}  
&Ref.~\cite{j}&Ref.~\cite{mlw}$^*$
&PDG~\cite{pdg}\\
\hline
$\Lambda_c$&$0(\frac12^+)$ & 2297&2265 &2285& & &2290  &2284.9(6)   \\
$\Sigma_c$&$1(\frac12^+)$ &2439&2440 &2453& & &2452  &2451.3(7)  \\
$\Sigma^*_c$&$1(\frac32^+)$ & 2518&2495 &2520&2518& & 2538 &2515.9(2.4) \\
$\Xi_c$&$\frac12(\frac12^+)$& 2481& &2468& & &2473  &2466.3(1.4)  \\
$\Xi'_c$&$\frac12(\frac12^+)$&2578& &2580&2579 & 2580.8(2.1)& 2599&
2574.1(3.3)\\ 
$\Xi^*_c$&$\frac12(\frac32^+)$&2654& &2650& & &2680 &2647.4(2.0)\\
$\Omega_c$ & $0(\frac12^+)$& 2698& &2710 & & &2678 &2697.5(2.6)\\
$\Omega^*_c$ & $0(\frac32^+)$&2768& &2770& 2768 &2760.5(4.9) & 2752 &\\
$\Lambda_b$&$0(\frac12^+)$& 5622& 5585&5620& & &5672  & 5624(9)   \\
$\Sigma_b$&$1(\frac12^+)$ &5805&5795 &5820& & 5824.2(9.0)& 5847 &  \\
$\Sigma^*_b$&$1(\frac32^+)$ & 5834&5805 &5850&  &5840.0(8.8)&  5871&\\
$\Xi_b$&$\frac12(\frac12^+)$& 5812&&5810& &5805.7(8.1)& 5788&\\
$\Xi'_b$&$\frac12(\frac12^+)$&5937&&5950& &5950.9(8.5)& 5936&\\
$\Xi^*_b$&$\frac12(\frac32^+)$&5963&&5980& &5966.1(8.3)& 5959&\\
$\Omega_b$ & $0(\frac12^+)$&6065&&6060& &6068.7(11.1)& 6040&\\
$\Omega^*_b$ & $0(\frac32^+)$ & 6088&&6090& &6083.2(11.0)& 6060& \\
\end{tabular}
\end{ruledtabular}
\flushleft{${}^*$ error estimates are about 50 MeV for charmed baryons and
  100 MeV for bottom baryons}
\end{table}

In Ref.~\cite{ci} the baryon masses are calculated in the framework of
a relativized quark model, applying a variational approach to obtain
the mass eigenvalues and bound state wave functions by using a harmonic
oscillator basis.  
In Ref.~\cite{rlp} the Feynman-Hellman theorem and semiempirical mass
formulas are used to predict the masses of heavy baryons. The
heavy-quark symmetry ($1/m_Q$ expansion) and $SU(3)$ flavour symmetry
are applied in Refs.~\cite{sav,ros,j} to evaluate the masses of
baryons with a single heavy quark. At lowest order in $SU(3)$ breaking
these masses obey an equal-spacing rule:

\begin{eqnarray}
  \label{eq:esr}
  J=\frac12, \qquad &&M_{\Sigma_Q}+M_{\Omega_Q}=2M_{\Xi'_Q},\cr\cr
J=\frac32, \qquad &&M_{\Sigma^*_Q}+M_{\Omega^*_Q}=2M_{\Xi^*_Q}, \quad
Q=b,c.
 \end{eqnarray}
 The corrections to this rule,
estimated on the basis of chiral perturbation theory (light meson
loops) combined with heavy-quark symmetry, are found to be small \cite{sav}.
The equal-spacing rule holds also for the hyperfine mass splittings
\cite{sav}:
\begin{eqnarray}
  \label{eq:hsms}
&& \delta_{\Sigma_Q}+\delta_{\Omega_Q}=2\delta_{\Xi_Q}, \qquad Q=b,c;\\\cr
  \delta_{\Sigma_Q}=M_{\Sigma^*_Q}-M_{\Sigma_Q};&& \qquad
\delta_{\Xi_Q}=M_{\Xi^*_Q}-M_{\Xi'_Q}; \qquad
\delta_{\Omega_Q}=M_{\Omega^*_Q}-M_{\Omega_Q}.\nonumber
\end{eqnarray}
This relation is expected \cite{j} to be more accurate than the relation
(\ref{eq:esr}).  

The hyperfine splitting calculation is used in \cite{ros} to estimate
the masses $M_{\Sigma^*_c}=2514$ MeV and $M_{\Omega^*_c}=2771$ MeV.
The heavy-quark expansion and broken $SU(3)$ symmetry are combined in
Ref.~\cite{j} with the $1/N_c$ expansion. As a result, some new mass
relations are obtained, which allowed to predict accurately the heavy
baryon masses. The accuracy of the mass relation (\ref{eq:hsms}) is
estimated there to be of order 1 MeV for $Q=c$ and 0.3 MeV for $Q=b$.

In Ref.~\cite{mlw} the masses of charmed and bottom baryons are
computed within quenched lattice nonrelativistic QCD (NRQCQ). The masses of
baryons with $b$ quark are also calculated in lattice NRQCD in
Ref.~\cite{ali}. The error bars of lattice calculations are usually of
order 50--100 MeV at present. 

From Tables~\ref{tab:tes} and  \ref{tab:t} it is evident that the
values of baryon masses obtained in the present paper (see
Table~\ref{tab:bm}) satisfy rather well both the mass relations
(\ref{eq:esr}) and (\ref{eq:hsms}). Note that these masses satisfy mass
inequality $2M_{\Xi_Q}\ge M_{\Sigma_Q}+M_{\Omega_Q}$ found from
analysis of the spectral properties of the Hamiltonians in
Refs.~\cite{lieb}.
 This gives a strong additional
support to our model, since it means that the model incorporates the
important features of broken $SU(3)$ flavour symmetry and heavy quark
expansion of QCD (see also \cite{fg}) in a reasonable way.

\begin{table}
\caption{\label{tab:tes} Test of validity of the equal-spacing rule
  (\ref{eq:esr})  
for heavy baryon masses obtained in this paper (in MeV).}
\begin{ruledtabular}
\begin{tabular}{ccccc}
& \multicolumn{2}{l}{\underline{\hspace{1.8cm}$J=\frac12$\hspace{1.8cm}}} 
\hspace{-2.1cm}
&\multicolumn{2}{l}{\underline{\hspace{1.8cm}$J=\frac32$\hspace{1.8cm}}}
\hspace{-2.1cm}\\
& $Q=c$ & $Q=b$ & $Q=c$ & $Q=b$\\
\hline
$M_{\Sigma_Q}+M_{\Omega_Q}$ & 5137 & 11870 & 5286 & 11922\\
$2M_{\Xi_Q}$ & 5156 & 11874 & 5308 & 11926
\end{tabular}
\end{ruledtabular}
\end{table}

\begin{table}
\caption{\label{tab:t} Test of validity of the equal-spacing rule
  (\ref{eq:hsms})  
for hyperfine mass splittings obtained in this paper (in MeV).}
\begin{ruledtabular}
\begin{tabular}{ccc}
 & $Q=c$ & $Q=b$\\
\hline
$\delta_{\Sigma_Q}+\delta_{\Omega_Q}$ & 149 & 52\\
$2\delta_{\Xi_Q}$ & 152 & 52 
\end{tabular}
\end{ruledtabular}
\end{table}

\section{Conclusions}
\label{sec:c}

It is important to emphasize that, in calculating the heavy baryon
masses, we do not use any free adjustable parameters. Indeed, all
parameters of the model (including quark masses and parameters of
the quark potential) have fixed values which were determined from
our previous considerations of  heavy and light meson properties. Note
that the light diquark in our approach is not considered as a
point-like object. Instead we use its wave functions to calculate  
diquark-gluon interaction form factors and, thus, take into account
the finite (and relatively large) 
size of the light diquark. The other important advantage of our
model is the completely relativistic treatment of the light quarks
in the diquark and the light diquark in the heavy baryon. We use the $v/c$
expansion only for heavy ($b$ and $c$) quarks.     
The overall reasonable agreement of our model predictions given in
Table~\ref{tab:bm} with both available experimental data and the results of
significantly
distinct theoretical approaches gives further grounds for the
heavy-quark--light-diquark picture of  heavy baryons.

The authors are grateful to A. Ali Khan, D. Antonov,  M. M\"uller-Preussker
and V. Savrin  for support and discussions.  Two of us
(R.N.F. and V.O.G.)  were supported in part by the {\it Deutsche
Forschungsgemeinschaft} under contract Eb 139/2-3 and by the {\it Russian
Foundation for Basic Research} under Grant No.05-02-16243.

\end{document}